\documentclass{iaus}

\usepackage{graphicx}
\usepackage{psfrag}

\title[evolution of very massive stars\ldots] %% give here short title %%
{The formation and evolution of very massive stars in dense
stellar systems}

\author[Belkus et al.]   %% give here short author list %%
{Houria Belkus$^1$, Joris Van Bever$^2$ \break \and Dany Vanbeveren$^{1,3}$}

\affiliation{
  $^1$Astrophysical Institute, Vrije Universiteit Brussel,
      Pleinlaan 2, 1050 Brussel, Belgium  \break
 email: hbelkus@vub.ac.be, dvbevere@vub.ac.be \break
  $^2$Institute of Computational Astrophysics, Saint Mary's University, \break
      923 Roby Street, Halifax, Nova Scotia, Canada B3H 3C3 \break 
      email: vanbever@ap.smu.ca  \break 
   $^3$ Mathematics department, Groep T, Vesaliusstraat 13, 3000 Leuven, Belgium           \\[\affilskip]

            }

\pubyear{2007}

\volume{246}  %% insert here IAU Symposium No.

\pagerange{119--126}

\date{?? and in revised form ??}

\jname{Proceedings Title IAU Symposium}

\editors{A.C. Editor, B.D. Editor \& C.E. Editor, eds.}

\newcommand{\msun}{\ensuremath{M_{\odot}}}	% solar mass symbol
\newcommand{\lsun}{\ensuremath{L_{\odot}}}	% solar luminosity symbol
\newcommand{\Teff}{\ensuremath{T_{\rm{eff}}}}
\newcommand{\TeffkK}[1]{#1\,\rm{kK}}

\begin{document}

\maketitle

\begin{abstract}

The early evolution of dense stellar systems is governed by massive single star
and binary evolution. Core collapse of dense massive star clusters can lead to the formation
of very massive objects through stellar collisions ($M\geq$ 1000\,\msun). Stellar wind mass
loss determines the evolution and final fate of these objects, and decides upon whether they
form black holes (with stellar or intermediate mass) or explode as pair instability
supernovae, leaving no remnant. We present a computationaly inexpensive evolutionary scheme
for very massive stars that can readily be implemented in an N-body code. Using our new
N-body code 'Youngbody' which includes a detailed treatment of massive stars as well as this
new scheme for very massive stars, we discuss the formation of intermediate mass and stellar
mass black holes in young starburst regions. A more detailed account of these results can be
found in \cite{Belkusetal}.



\end{abstract}

Our calculations of the evolution of very massive stars are based on the
results of \cite{Nadyozhin} who constructed stellar structure models for these
objects using the similarity theory of stellar structure. Their models correspond
to chemically homogeneous stars, having Thompson scattering as the only opacity
source throughout. This provides an accurate treatment, since extremely massive
stars are almost completely convective during their evolution, and the opacity
differs significantly from the Thompson scattering value only in a thin layer
near the surface of the star. Our stellar parameters should therefore be reliable,
with the possible exception of the effective temperature.

\cite{Kudritzki} studied line-driven stellar winds of very massive stars and
calculated mass loss rates of very massive O-type stars as a function of
metallicity,  in the range $6.3 \leq \log(L/\lsun) \leq 7.03$ and $\TeffkK{40}
\leq \Teff \leq \TeffkK{60}$.  The mass loss rates are smallest for the highest
\Teff. We use his interpolation formula for the
\TeffkK{40} models to compute the mass loss rates of our evolutionary models,
meaning that we obtain an upper limit to the mass  of a model star at all times.
Note that for a given luminosity, we ensure that the mass loss rate never
exceeds the maximum for line-driven winds, as given by \cite{Owocki}.

\begin{figure}[ht]

\begin{minipage}[t]{0.5\linewidth}

\centering
\psfrag{Y}{{\tiny $M_{\textrm{end}}(\msun)$}}
\psfrag{Z}{{\tiny$M_{\textrm{init}}(\msun)$}}
\includegraphics[scale=0.5]{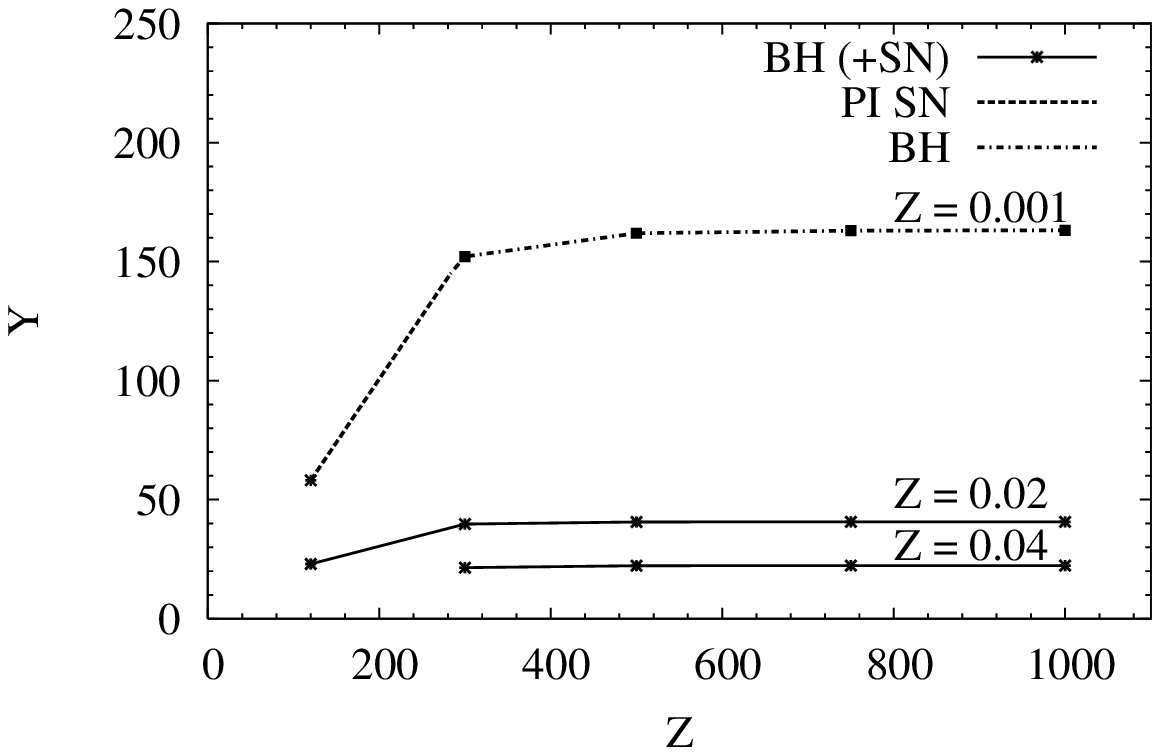}
 \caption{Masses of massive and very massive stars ($X_{\rm{c,0}}=0.68$) at the
end of the core helium burning stage for 3 different metallicities (see labels).
Note the almost constant final mass of very massive stars, which is due to the
properties of the $\dot{M}-L$ relationship that is used. Also, the dashed part of the
$Z=0.001$ curve indicates stars that are expected to explode as pair instability
supernovae, leaving no remnant.}
\label{fig:final_masses}

\end{minipage}
\hspace{0.5cm}
\begin{minipage}[t]{0.5\linewidth}

\centering
\psfrag{M}{{\tiny M (\msun)}}
\includegraphics[scale=0.5]{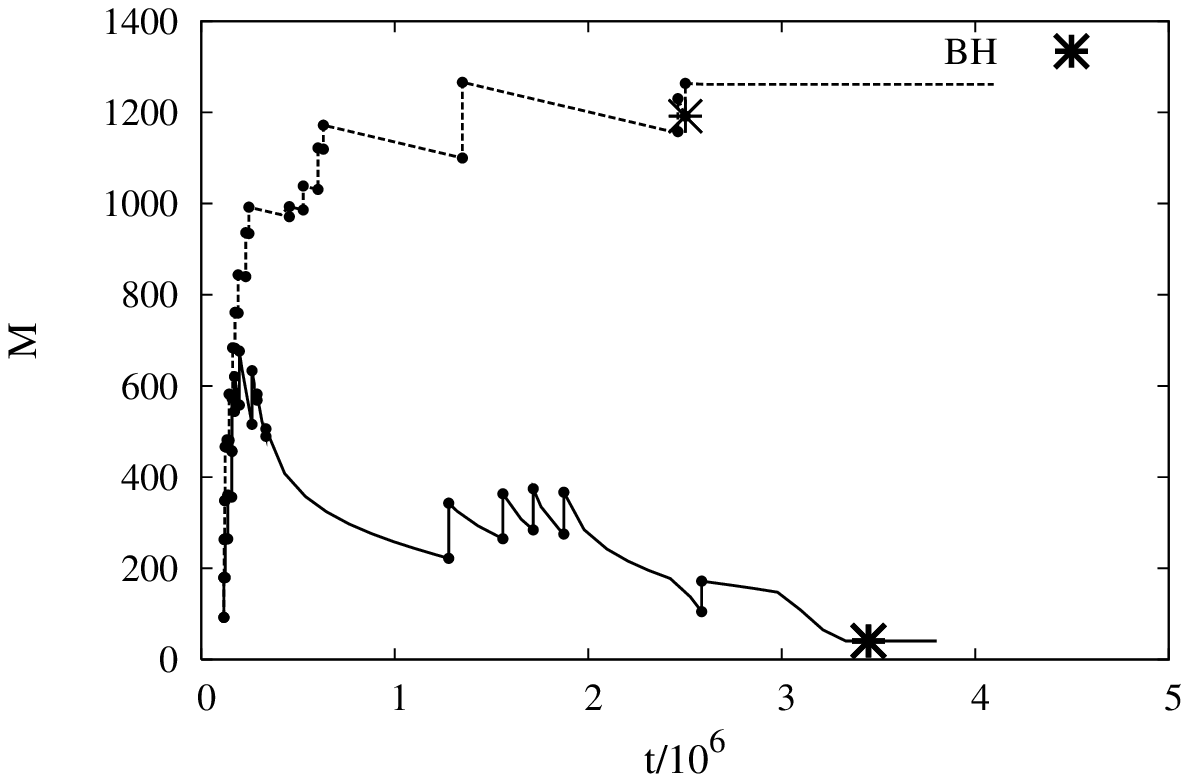}
 \caption{Evolution of the mass of the runaway merger in a King ($W_0$ = 9) N-body
model containing 3000 massive single stars ($M\geq 10\,\msun$) and with a half mass
radius of 0.5 pc. The doted line represents a model in which any star more massive than 120 M0 is given a constant mass loss rate of $10^{-4} \msun/yr$. The full line represents a model in which those same stars are treated with the mass loss rates of Kudritzki (2002). The star symbols denote the moment at which the runaway merger collapses into a black hole.}
\label{fig:runaway}

\end{minipage}

\end{figure}

Figure~\ref{fig:final_masses} shows the masses of very massive stars 
(with initial central hydrogen abundance $X_\textrm{c,0}=0.68$) at the end of the core helium burning stage for 3 different
metallicities. It is seen that only very massive stars at sufficiently low
metallicities are expected to produce pair-instability supernovae and direct collapse black holes (compared to the limiting masses from \cite{Hegeretal}). At
high $Z$, the stellar wind mass loss rates reduce the stellar mass to such an
extent that only black holes due to accompanying supernova and fallback result.

We implemented this evolution scheme in our direct N-body code 'Youngbody' and
computed models of young dense starburst regions showing the creation of so-called
runaway mergers. Figure~\ref{fig:runaway} shows a typical example and indicates
that the reduction of mass by stellar wind is able to compete with the growth of the
runaway star due to stellar collisions, at least in the later stages. In this model
the core collapse stage is over before the merger ends its life and therefore mass
loss is able to reduce the stellar mass sufficiently to prevent the formation of an
intermediate mass black hole. This suggests that Ultra Luminous X-ray sources (ULXs) may not
be IMBH accreting at rates close to the Eddington rate, but could be stellar mass BHs
($\approx 50-100\,\msun$) accreting at Super Eddington rates (\cite[Soria 2007]{Soria}).

\end{document}